%% file: main.tex
\documentclass[a4paper]{article}

\usepackage{a4wide}
\usepackage{lineno,hyperref}
\modulolinenumbers[5]
\hypersetup{
    colorlinks=true,
    linkcolor=blue,
    filecolor=magenta,
    citecolor=blue,  
    urlcolor=blue,
    pdftitle={Estimating the number of COVID-19 cases from population isolation level},
    pdfpagemode=FullScreen,
}
\usepackage{amsmath}
\usepackage{amssymb}

\input macros
\def\eq#1{Equation~#1}
\def\fig#1{Figure~#1}

\begin{document}

\title{Estimating the number of COVID-19 cases from population isolation level}
\author{\href{https://dcm.ffclrp.usp.br/~aholanda}{Adriano J.\ Holanda}$^{1,2}$\footnote{Contact: \href{mailto:aholanda@usp.br}{aholanda@usp.br}}\\
\small $^1$Department of Computing and Mathematics\\
\small Faculty of Philosophy, Science and Letters at Ribeir\~{a}o Preto \\
\small University of S\~{a}o Paulo \\
\small \hbox{3900 Bandeirantes Ave.},Ribeirão Preto, S\~{a}o Paulo, 14040-901, Brazil\\
\small $^2$Faculty ``Dr.\ Francisco Maeda'', Ituverava, S\~{a}o Paulo, Brazil}
\maketitle


\begin{abstract}
We use the logistic function to estimate the number 
of individuals infected by a virus in a period of time
as a function of social isolation level 
in the previous period of the infection occurrences.
Each period is composed by a fixed date range in days 
which the social isolation is supposed to take
effect over the virus spread in the next 
date range.
The sample is the COVID-19 cases and 
social isolation level data from S\~{a}o Paulo State, Brazil.  
The proposed method is divided into two stages: 
1) The logistic function is fitted against COVID-19 
empirical data to obtain the function parameters; 
2) the function parameters, 
 except for the overall growth rate,
 and the mean of social isolation level for all periods of time 
are used to calculate a constant called $\lambda$.
The logistic growth rate for each period of time
is calculated using $\lambda$ 
and the isolation level for that period.
 The number of cases in a period is estimated using 
the logistic function and the growth rate 
from previous period of time 
to obtain the effect of social isolation during the
elapsed time. 
The period of time that produces a 
better correlation between empirical and estimated 
data was $5$ days. 
We conclude the method performs a data estimation with  
 high correlation with the empirical data.
\end{abstract}



\section{Introduction}
\label{intro}

SARS-CoV-2 is the virus that causes the infection 
called coronavirus disease (COVID-19). 
COVID-19 pandemic control will depend 
on the coverage of population immunity obtained 
through infection or vaccination~\cite{WHOa}. 
The mutants of SARS-CoV-2 impose a 
predictability challenge in the long-term efficacy 
of the current vaccines. 
 It's of utmost importance the preparation of updated 
 vaccines tailored to emerging variants that are cross-reactive 
 against all circulating variants~\cite{Harvey2021}. 
Indeed, the percentage of immunized population in some countries 
is not enough to shutdown the SARS-CoV-2 spread, 
e.g., 54.3\% of the population  
is fully vaccinated in Brazil~\cite{MH2021}.
The SARS-CoV-2 variants and its low rate of immunization 
impose the continuation of non-pharmacological approaches: 
handwashing, use of facial masks and social distancing~\cite{WHOb}. 
Among these approaches, only social distancing 
can be assessed effectively using 
communication traces~\cite{Farrahi2014}. 
 Isolating cases, quarantining contacts and implementing
large-scale social distancing approaches have been proved to 
be effective in the control of virus spread~\cite{Aquino2020}.

In this study, we use the logistic function 
to calculate the expected number of COVID-19 cases 
as a function of time
based 
on the isolation level of the population and 
observed number of cases. 
We've chosen the S\~{a}o Paulo State, Brazil data as sample, 
but the method is simple and generic 
to apply to any population data. 
S\~{a}o Paulo is one and the most populous of the 26 states 
of the Federative Republic of Brazil with 
more than 40-million inhabitants, 21.9\% 
of Brazilian population and is responsible for 33.9\% 
of Brazil's GPD (Gross Domestic Product) 
(\url{https://www.ibge.gov.br/cidades-e-estados/sp.html}).

\section{Methods}
\label{methods}

\subsection{Data preparation}

We gathered the number of cases of COVID-19 
in S\~{a}o Paulo State, Brazil, 
from Github repository of ``S\~{a}o Paulo Plan'' 
(\url{https://github.com/seade-R/dados-covid-sp}) 
that is a initiative from the state government 
to organize and process data related to COVID-19
from all cities of the state. 
The plan aims to help in the decision-making 
along the application of public 
procedures to control the SARS-CoV-2 spread. 
We also fetched the social isolation level data 
for all state cities  from 
``S\~{a}o Paulo Plan'', but from another web address 
(\url{https://www.saopaulo.sp.gov.br/coronavirus/isolamento/}).

In order to avoid large discrepancies 
in the number of cases due delayed notification, 
 each data point is composed by a group of days 
 which each group is labeled by the most 
 recent date in the group and 
 all groups have the same number of days. 
 All data points contain 
 the sum of the COVID-19 cases and 
 the mean of the population isolation levels 
 in the group for all cities 
 located in the S\~{a}o Paulo State.

\subsection{Fitting}

We choose the logistic function to model 
the SARS-CoV-2 spread behavior. We use the 
solution for

\begin{equation}
\frac{dN}{dt} = r N (1 - \frac{N}{K}),
\label{eq:df:logistic}
\end{equation}

\noindent that is

\begin{equation}
	N(t) = \frac{K}{1- \frac{(K-N_{m})}{N_{m}} e^{-rt}},
\label{eq:exp:logistic}
\end{equation}

\noindent where $K$ is the maximum value reachable by $N$, 
$N_m$ is the curve midpoint, 
$r$ is the growth rate 
and $t$ is the time.

\begin{figure}
\centering
\includegraphics[scale=.4725]{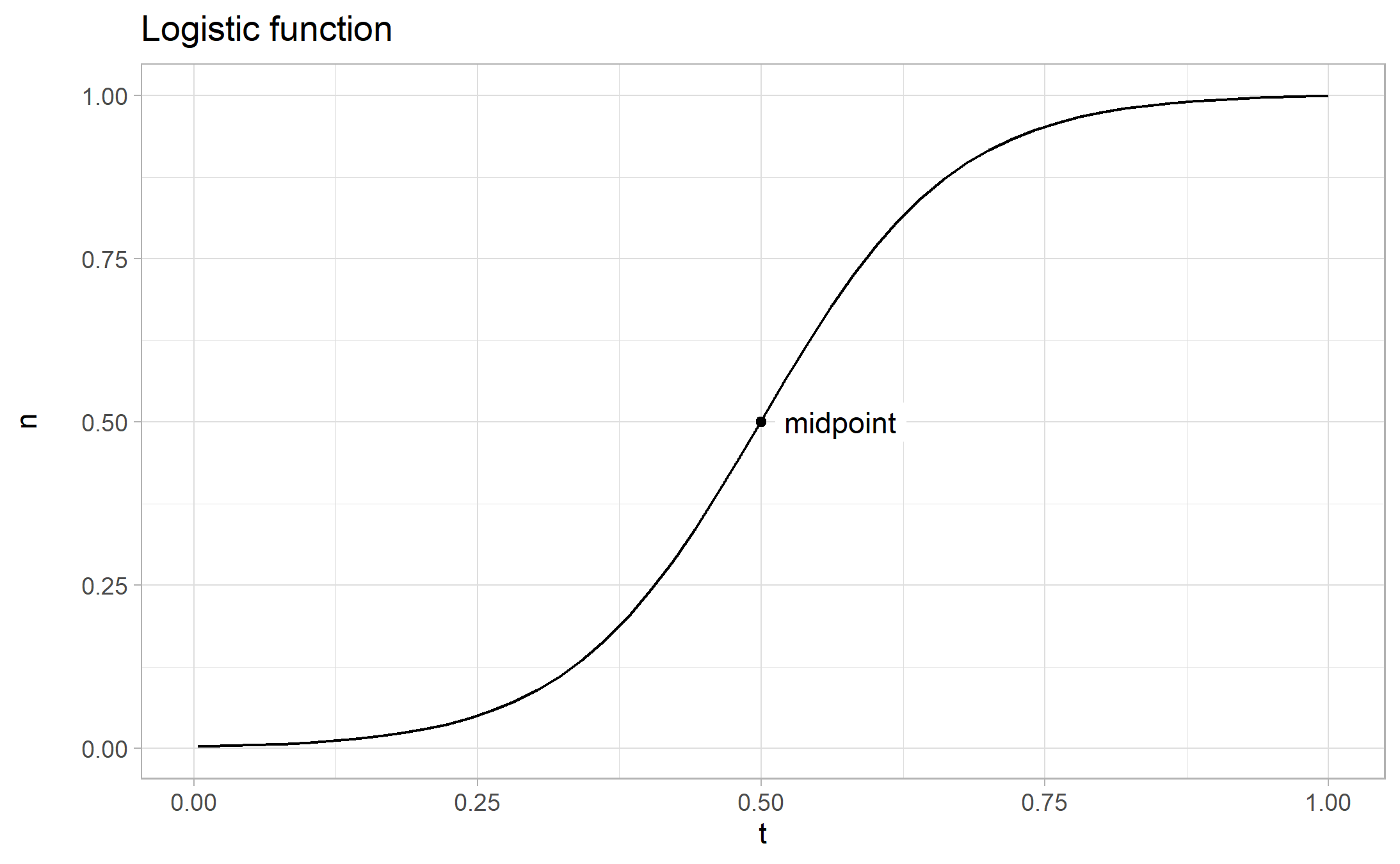}
\caption{Hypothetical normalized logistic function -- 
curve of \eq{\ref{eq:df:logistic}} with 
$K=1.0$ and an arbitrary growth rate $r=0.04$; 
the midpoint coordinates are $(0.5, 0.5)$.
The $K$ represents a normalized measure where 
the maximum value is the divisor factor.}
\label{fig:logistic}
\end{figure}

\fig{\ref{fig:logistic}} shows a hypothetical logistic function 
plotted using \eq{\ref{eq:df:logistic}} where before the midpoint 
the number of cases between 
two consecutive time steps increases, 
and after the midpoint decreases. 
The curve reaches a plateau in the maximum value for $n(t)$. 
The number of cases $n$ and time $t$ are normalized.
In an empirical scenario, the plateau could be achieved 
if the process that is feeding the curve 
growth is controlled with the maximum 
value for $n(t)$ not being reached.

We considered $N$ to be the number of infected individuals 
and choose the logistic function because of its behavior, 
with an exponential increase at the beginning but slowing down after 
a  midpoint and reaching a plateau. 
The same behavior occurs with a virus dissemination when at the 
beginning it spreads exponentially, and it starts to 
decrease the contamination after some control protocol is implemented, 
like vaccination or lockdown, 
or in the worst case scenario when the majority of the population is infected.

The graph of the cumulative number of empirical cases of COVID-19 
at each time step was plotted, 
and the \eq{\ref{eq:exp:logistic}} was fitted against the 
empirical data.

\subsection{Estimation}

We postulate that the growth rate $r$ in the logistic function is 
inversely proportional to the population isolation level $i$ 
modulated by a constant $\lambda$:

\begin{equation}
r_{\text{fit}} = {\lambda /\, \overline{i\,}} \Rightarrow \lambda = r_{\text{fit}}\, \overline{i\,}.
\label{eq:lambda}
\end{equation}

We calculate the constant by multiplying 
the mean isolation level $\overline{i\,}$  
of all time steps by 
$r_{\text{fit}}$ obtained with the fitting of 
logistic function and the empirical data.

The constant calculated using \eq{\ref{eq:lambda}}
and isolation level of each time interval $i(t)$ 
are used to estimate the growth rate $r(t)$ using the formula:

\begin{equation}
r(t) = \frac{\lambda}{i(t)}.
\label{eq:rt}
\end{equation}

A time step is a period of time in days used to group the 
COVID-19 cases where the social isolation level 
from the last period takes effect 
 over the SARS-CoV-2 spread. 
The estimated number of cases $N(t)$ is calculated using
 \eq{\ref{eq:exp:logistic}} plugging $r(t-1)$ instead of $r(t)$ because 
 we take the previous growth rate $r(t-1)$ 
 to predict the effect in the number of cases 
 after a period of time $N(t)$:

\begin{equation}
N(t) = \frac{K}{1- \frac{(K-N_{m})}{N_{m}} e^{-r(t-1)\,t}}.
\label{eq:exp:logistic:estimation}
\end{equation}

 We compare these results with the empirical data using Pearson correlation.
 We use a time step $t$ 
 equals to $\pgfkeysvalueof{/pandiso/time/step}$ 
 days because it was the value with
 a better correlation between empirical and estimated data. 
Furthermore, we only use integers as time intervals for days because
 the variation in the correlation values was not
 significant to justify handling fractional 
values  that increases the complexity
 of code and analysis.

\section{Results}
\label{results}

\begin{figure}[ht]
\centering
\includegraphics[scale=.545]{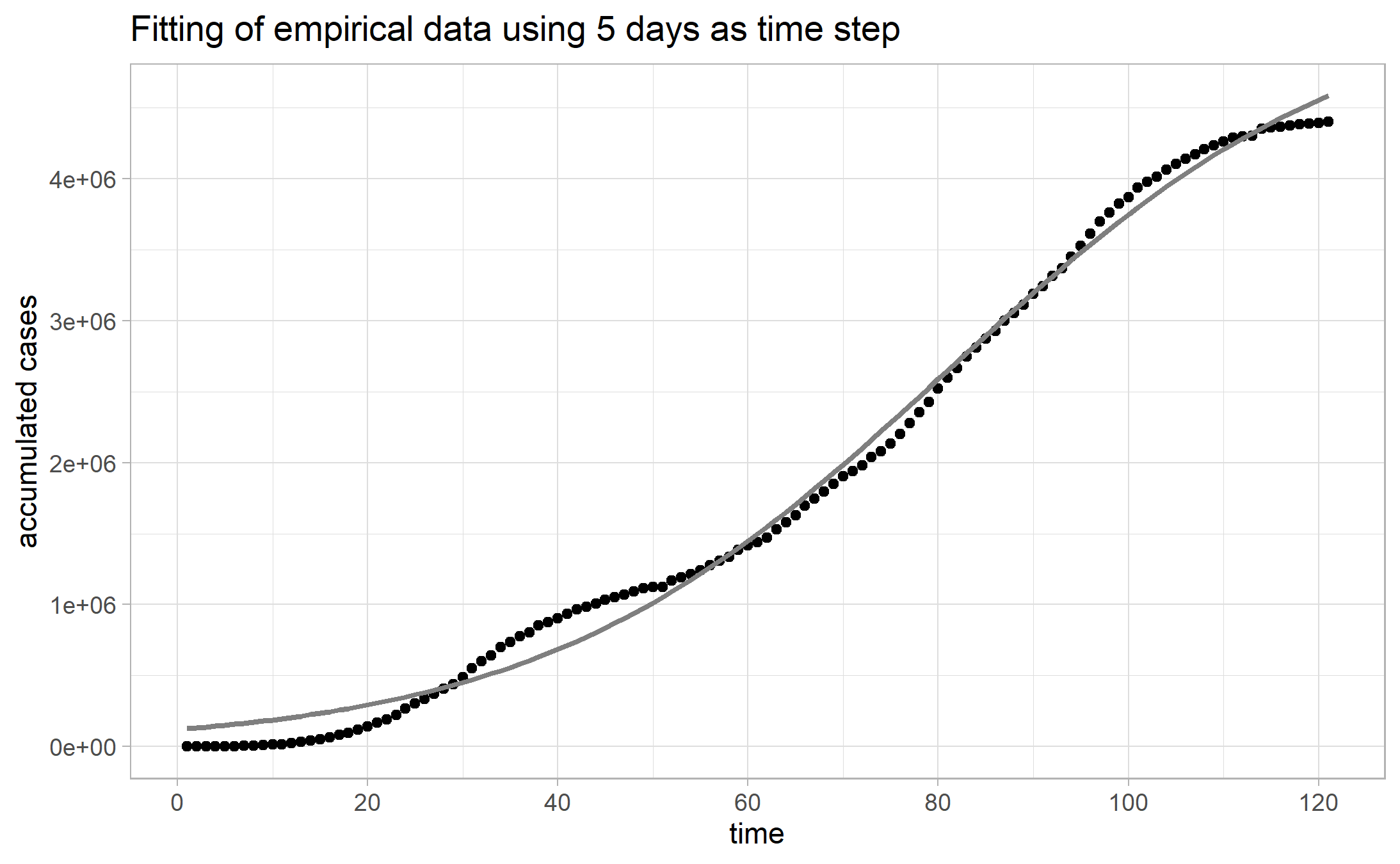}
\caption{Curve of accumulated number of cases 
at each time step $t$ and its fitting curve 
using the logistic function (\eq{\ref{eq:exp:logistic}}) 
where the parameters obtained from fitting are: 
$r=\pgfkeysvalueof{/pandiso/fit/r}$, 
$N_m=\pandisofitm$, 
$K=\pandisofitK$. 
The Pearson correlation between the empirical 
and fitted data is $\pgfkeysvalueof{/pandiso/fit/pearson}$ and 
 $\text{p-value} < 2.2\times 10^{-16}$.  
The  number of time steps is $\pgfkeysvalueof{/pandiso/count/time/steps}$
and the initial and final dates are ``\pgfkeysvalueof{/pandiso/date/first}''
and ``\pgfkeysvalueof{/pandiso/date/last}'', respectively.}
\label{fig:fit}
\end{figure}

\fig{\ref{fig:fit}} shows the accumulated 
number of COVID-19 cases per 
time step in São Paulo State and 
the fitting performed using logistic 
function~(\eq{~\ref{eq:exp:logistic:estimation}}) 
that proved to be a good approximation to the 
cumulative growth of number of cases.

We notice the consistent increase in the number of cases 
with the plateau to be achieved when the number of cases reaches 
$\pandisofitK$ 
according to the parameters extracted from fitting.

\begin{figure}[ht]
\centering
\includegraphics[scale=.55]{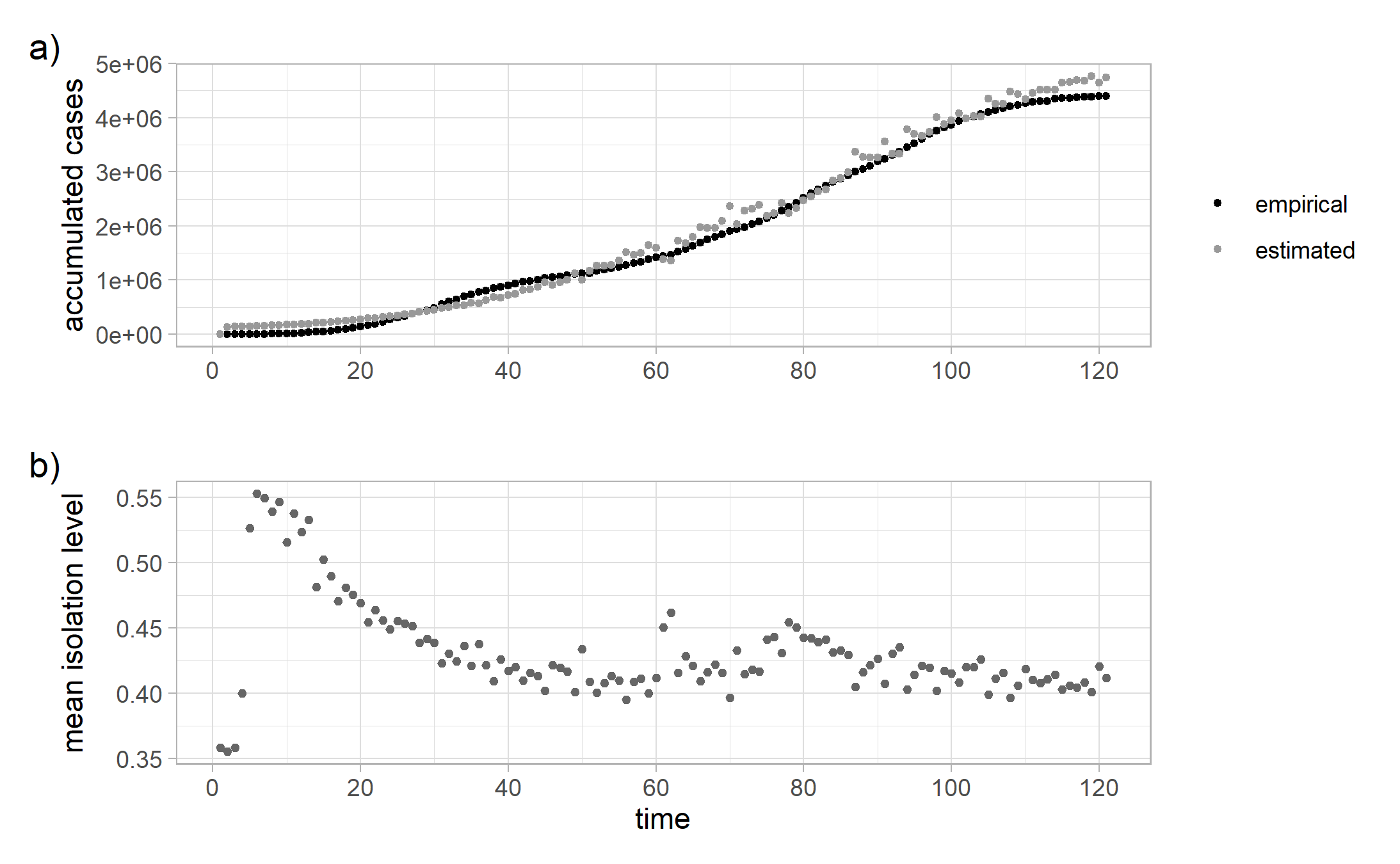}
\caption{a) Curve of accumulated number of cases 
and the estimation using the logistic 
function~(\eq{\ref{eq:exp:logistic:estimation}}) 
with the growth rate $r(t)$ obtained from \eq{\ref{eq:rt}}. 
The constant $\lambda$ was calculated using \eq{\ref{eq:lambda}} with 
$r=\pgfkeysvalueof{/pandiso/fit/r}$ 
and $\,\overline{i\,} = \pgfkeysvalueof{/pandiso/overall/level/mean} 
 \pm \pgfkeysvalueof{/pandiso/overall/level/stdev}$. 
The Pearson correlation coefficient between empirical and estimated data is 
$\pgfkeysvalueof{/pandiso/estimation/pearson}$ 
and $\text{p-value} < 2.2\times 10^{-16}$. 
b) Mean isolation level at each time step $t$. 
The number of steps, initial and final dates are 
the same as \fig{\ref{fig:fit}}.}
\label{fig:estimation}
\end{figure}

\fig{\ref{fig:estimation}} shows the empirical curve again 
and the estimated number of cases using 
the fitting parameters $K$ and $N_m$ 
and the calculation of growth rate $r(t)$ using \eq{\ref{eq:rt}}, 
where the isolation level $i(t)$ is used, 
to plug into logistic function (\eq{\ref{eq:exp:logistic:estimation}}) 
and calculate the estimated number of cases $N(t+1)$.

We noticed at the beginning of the curve in 
\fig{~\ref{fig:estimation}b}, high values of isolation levels. 
At that moment, the S\~{a}o Paulo State government 
adopted more strict rules to avoid circulation of people. 
It was issued from March 16th and March 22nd a decree 
to encourage home-based working whenever possible, 
closure of schools and non-emergencies commerce\cite{Cruz2020}.

\section{Discussion}
\label{discussion}

The proposed method to predict the number of infected individuals 
using the current empirical data and the logistic function proved 
to be an effective approximation for the infection spread. 
It's a simple method and the fact it doesn't take into account 
other barriers to the infection spread, like the use of a facial mask, 
is a good property since it's very difficult to obtain the correlation about 
the application of these fundamental procedures and the number of cases.

The main reason to develop a simple but accurate method to estimate 
the number of infected individuals using the isolation level is to help 
in the decision-making process related to the control of infection spread. 
The isolation level is a reliable measure and may be used to predict 
the desired behavior of the population in time steps and check if the infection 
is really being controlled, reaching a remission in an expected time ahead.

We adopted a simpler method instead of more detailed 
compartmental models~\cite{Ross1916,Ross1917a,Ross1917b}, 
e.g. SIR (Susceptible, Infectious, or Recovered), SEIR (Susceptible, Exposed, Infectious, or Recovered) 
or SEIS (Susceptible, Exposed, Infectious, then Susceptible again), due to the following reasons: 
a) The number of infected individuals is very small when compared with population if there is 
    an application of approaches to control the virus spread; 
b) It's difficult to assess the influence of non-pharmacological approaches and the approximation 
    used somewhat seems to embed this influence due to the high correlation 
   the empirical and estimated data; 
c) Births, deaths, deaths due to the disease, immigration and emigration rates 
   are also difficult to assess effectively in most countries, 
   even deaths due to the disease 
   may be contaminated by errors related to death certificate misjudgment; 
d) It's challenging to assess, in some of these models, the number of 
    individuals with repeated infections and the gradual loss of acquired 
    immunity because some aspects of the immunization process is 
   yet to be unraveled, at least for COVID-19.

Another advantage of our method over the compartmental models is the 
 reduction of dimensionality to one manageable and reliable dimension, 
this process is also called feature engineering~\cite{Spieg2019}. 
This property also facilitates the source code 
understanding, maintenance and efficiency. 
The data is easily updated and the number of days in each time step 
may be changed in the source code.

For future developments, we visualize the use of Control Theory 
in the estimation phase~\cite{Stewart2020}. 
The isolation level and number of cases 
may be used as process variable and set point, respectively, 
in a controller like PI (Proportional-Integral) or PID (PI-Derivative), 
optimizing the system according to some properties. 
The properties may be the vaccination rate and the vaccines’ stock, 
number of ICU (Intensive Care Unit) beds in a hospital, 
and even immunization loss growth rate.

\section*{Source code and data availability}
R code and data required to reproduce the analyses 
and Figures are available from
\url{https://github.com/aholanda/pandiso}.


\bibliographystyle{plain}
\bibliography{main}

\end{document}

%% file: macros.tex
\usepackage{tikz}
\def\pandisofitK{5,295,818}%
\def\pandisofitm{119,523}%
\pgfkeys{%
	/pandiso/count/time/steps/.initial=121,%
	/pandiso/date/first/.initial=2020-03-01,%
	/pandiso/date/last/.initial=2021-10-25,%
	/pandiso/estimation/pearson/.initial=0.996,%
	/pandiso/fit/pearson/.initial=0.997,%
	/pandiso/fit/r/.initial=0.047,%
	/pandiso/overall/level/mean/.initial=0.432,%
	/pandiso/overall/level/stdev/.initial=0.038,%
	/pandiso/time/step/.initial=5,%
}%